\documentclass[reprint,superscriptaddress,twocolumn,aps,prb,showpacs]{revtex4-1}
\usepackage{graphicx}
\usepackage{amsbsy,amssymb,amsmath,bm,ulem}

\graphicspath{{../fig/}}

\usepackage{color}

\newcommand{\removed}[1]{}

\begin{document}

\title{Dynamical barrier for flux penetration in a superconducting film in the flux flow state}

\author{J. I. Vestg{\aa}rden}
\affiliation{Department of Physics, University of Oslo, P. O. box
1048 Blindern, 0316 Oslo, Norway}
\author{Y. M. Galperin}
\affiliation{Department of Physics, University of Oslo, P. O. box
1048 Blindern, 0316 Oslo, Norway}
\affiliation{Ioffe Physical Technical Institute, 26 Polytekhnicheskaya, 
St Petersburg 194021, Russian Federation}
\author{T. H. Johansen}
\affiliation{Department of Physics, University of Oslo, P. O. box
1048 Blindern, 0316 Oslo, Norway}
\affiliation{Institute for Superconducting and Electronic Materials,
University of Wollongong, Northfields Avenue, Wollongong, NSW 2522,
Australia}

\begin{abstract}
The penetration of transverse magnetic flux into a thin
superconducting square film in the flux flow state is considered by
numerical simulation. Due to the film self-field, the governing
equations are nonlinear, and in combination with the finite viscosity
of the moving vortices, this sets up a dynamical barrier for flux
penetration into the sample. The corresponding magnetization loop is
hysteric, with the peak in magnetization shifted from the zero
position.  The magnetic field in increasing applied field is found to
form a well-defined front of propagation. Numerical estimates shows
that the dynamical barrier should be measurable on films with low volume
pinning.
\end{abstract}

\pacs{74.25.Ha,  74.78.-w }

\maketitle

\section{Introduction}
The penetration of magnetic flux into superconductors is delayed due
to the presence of surface barriers, such as the Bean-Livingston
barrier,\cite{beanlivingston64,burlachkov91,konczykowski91,olsen04} 
surface pinning\cite{flippen95}, and various
barriers of geometric origin.\cite{clem73,brandt93-epl,mawatari03}  
(The review by Brandt\cite{brandt95-rpp} lists 7
different mechanisms) The barriers are particularly
important in thin films where the equilibrium field for existence of
magnetic flux is much reduced from the bulk lower critical field $H_{c1}$
to $H_{c1}d/2w$, where $d$ is thickness and $w$ is sample width.\cite{zeldov94-prl} 
The presence of surface barriers implies that vortices will not necessarily enter 
the sample when it is energetically favorable for them to reside in the sample center.
Of particular importance in thin films, is the geometric barrier
caused by the magnetic fields piling up near the edges, which delays penetration
until the external field reaches $H_{c1}\sqrt{d/w}$.\cite{zeldov94-prl}
Numerical simulations show that in samples without volume pinning, the
magnetic flux that overcomes the barrier tends to pile up in the
sample center.\cite{brandt99} Because the barrier does not prevent
vortices from leaving the sample, the magnetization loop is
asymmetric, and the magnetization irreversible.\cite{brandt99-2}

The attention so far has mainly been paid to the static nature of
barriers. Yet, dynamic effects might also give rise to barriers for 
flux penetration. In order to investigate if this is the case we consider 
dynamics of a superconducting film in
transverse magnetic field. We assume that the film is sufficiently
wide, so that the magnetic field can be treated as a continuum, and
the spatio-temporal evolution of the system can be obtained by
solution of the Maxwell-equations. In order to separate the dynamical barrier from
other kinds of surface barriers, we disregard surface pinning, and assume that $H_{c1}=0$
and the critical current density, $j_c$,
is zero. Then, the only mechanism that gives loss in the system is the
finite viscosity of the moving vortices, which gives a flux flow
resistivity $\rho=\rho_n|H_z|/H_{c2}$, where $H_{c2}$ is the upper
critical field.  The corresponding dynamical barrier towards flux
penetration will thus be strongly dependent on the rate of change the
applied field.

\section{Model}
Let us consider a thin superconducting film with thickness $d$, 
shaped as a square with sides $2a\gg d$. Due to absence of pinning, 
$j_c=0$, and the resistivity is solely given by the conventional 
flux flow expression\cite{bardeen65}
\begin{equation}
  \rho=\rho_n|H_z|/H_{c2},
  \label{rho}
\end{equation}
where $\rho_n$ is the normal state resistivity, 
$H_{c2}$ is the upper critical field,
and  $H_z$ is the transverse component of the magnetic field.
The magnetic field has two contributions, the applied field 
and self-field of the sample,\cite{vestgarden13-fftsim}
\begin{equation}
  \label{h1}
  H_z=H_a+\mathcal F^{-1}\left[\frac{k}{2}\mathcal F\left[g\right]\right]
  ,
\end{equation}
where $\mathcal F$, and $\mathcal F^{-1}$ are forward and inverse
Fourier transforms respectively, and $k=\sqrt{k_x^2+k_y^2}$ is the
wave-vector. 
The local magnetization $g$ is defined by
$\nabla\times\hat zg=\mathbf J$, where $\mathbf J$ is the sheet
current.  The inverse of Eq.~\eqref{h1} and a time derivative gives
\begin{equation}
  \label{dotg1}
  \dot g 
  = 
  \mathcal F^{-1}\left[\frac{2}{k}\mathcal F\left[\dot H_z-\dot H_a\right]\right]
  .
\end{equation}
Inside the sample, Faraday law  and the 
material law, Eq.~\eqref{rho},  gives 
\begin{equation}
  \dot H_z=\nabla\cdot \left(H_z\nabla g\right) \rho_n /(H_{c2}\mu_0),
\end{equation}
where $H_z$ is given from Eq.~\eqref{h1}. Outside the sample, $\dot H_z $ 
is calculated by an iterative 
Fourier space -real space hybrid method which ensures $g=0$ in 
the vacuum outside the sample. \cite{vestgarden13-fftsim}
Eq.~\eqref{dotg1} is non-linear due to the self-field of the sample. 
In this respect, the situation is different from the parallel geometry,
where only the constant applied field enters the expression, and the  
corresponding equation for the flux dynamics is linear. 

Let us rewrite the equations on dimensionless form, assuming that the applied field 
is ramped with constant rate $|\dot H_a|$.
We define a time scale and sheet current scale as 
\begin{equation}
  t_0\equiv \sqrt{\frac{\mu_0H_{c2}dw}{\rho_n|\dot H_a|}},\qquad
  J_0\equiv \sqrt{\frac{\mu_0H_{c2}dw|\dot H_a|}{\rho_n}}
  .
\end{equation}
The dimensionless quantities are defined as 
$\tilde t = t/t_0$, $g/J_0w$, $\tilde H=H/J_0$, $\tilde k=wk$. Eq.~\eqref{dotg1}
becomes
\begin{equation}  
  \label{dotg2}
  \frac{\partial \tilde g}{\partial \tilde t} 
  = \mathcal F^{-1}\left[\frac{2}{\tilde k}\mathcal F
    \left[
      \frac{\partial \tilde H_z}{\partial \tilde t} -
      1
      \right]\right]
  ,
\end{equation}
where 
\begin{equation}
  \frac{\partial \tilde H_z}{\partial \tilde t}
  =
  \tilde \nabla \cdot \left[\tilde H_z\tilde \nabla \tilde g\right]
  ,
\end{equation}
valid inside the sample. As long as $|\dot H_a|$ is constant, 
there are no free parameters in the problem.  We will
henceforth omit the tildes in the dimensionless quantities,
when reporting the results.

A total are of size $1.4\times 1.4$ is discretized on a $512\times
512$ grid. The additional vacuum at the sides of the superconductor is
used to implement the boundary conditions.

\begin{figure}[t]
  \centering
  \includegraphics[width=6cm]{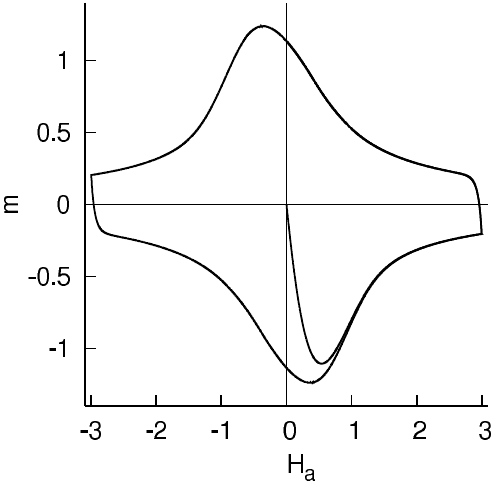} \\
  \caption{
    \label{fig:moment}
    The $m - H_a$ magnetization loop. Even in absence of pinning 
    the loop is hysteric due to the dynamical barrier.
  }
\end{figure}

\section{Result}
Let us now consider the evolution of the sample as it completes a 
magnetization cycle. The external field driven with 
constant rate $|\dot H_a|=1$ until the maximum field $H_a=3$, starting
from zero-field-cooled conditions. As applied field is changed, 
shielding currents are induced in the sample,  
giving it a nonzero magnetic moment $m$. The magnetic moment is
calculated as $m=\int g(x,y) dxdy $. Figure~\ref{fig:moment} shows
the magnetic moment as a function of applied field. The plot contains the 
virgin branch and a steady state loop. As expected for a superconducting 
film, the main direction of the response is
diamagnetic. The absolute value of $|m|$ reaches a peak for  $H_a=0.54$
in the virgin branch and at  $H_a=0.35$ in the steady-state loop, while 
it decreases at higher magnetic fields. The shape of the loop is quite similar to 
superconductors with a field-dependent critical current,\cite{mcdonald96}
except that the magnetization peak is shifted from $H_a=0$.\cite{shantsev99}
In this respect the dynamical barrier is similar to other kinds of 
surface barriers.\cite{burlachkov91,mawatari03}

\begin{figure*}[tbp]
  \centering
  \includegraphics[width=17cm]{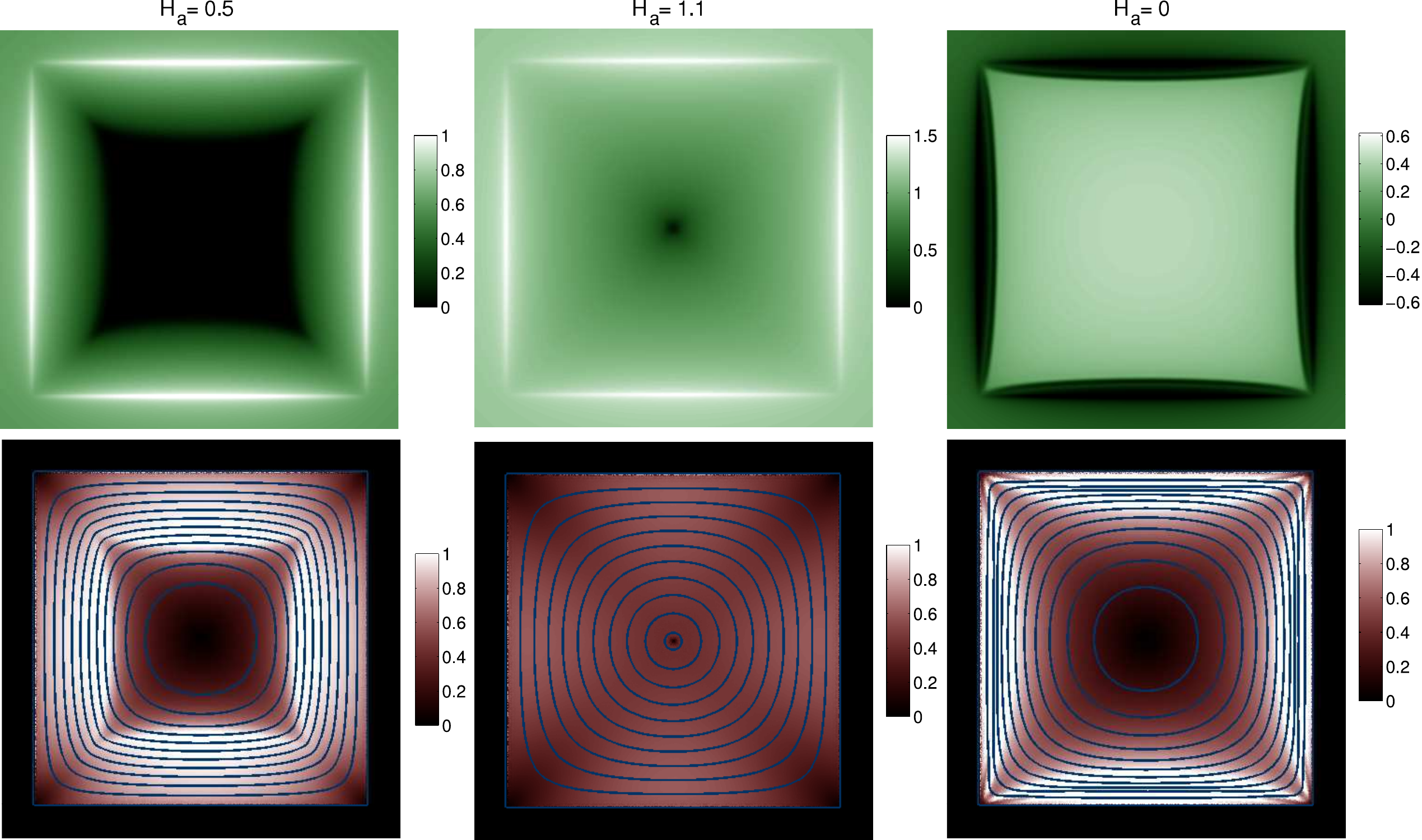} 
  \caption{
    \label{fig:HJ}
    (top) The magnetic flux distribution and (bottom) current density and stream lines, 
    at $H_a=0.5$, $1.1$, and $0$ (Remanent state).
  }
\end{figure*}

Figure \ref{fig:HJ} shows $H_z$ and $J$ magnitude and stream lines at
various applied fields. 
The state at $H_a=0.5$ is close to the peak in
magnetization in the virgin branch. The flux piles up close to the
edges, and falls to zero on a well defined flux front, roughly
penetrating one third of the distance to the sample center. The
current stream lines are smooth, with highest density in the 
flux-penetrated region. 
The flux distribution has some similarity with the square in the critical state,\cite{brandt95} 
but the most striking difference is the absence of dark $d$-lines at the diagonals.
At $H_a=1.1$, the flux front has reached the center of the sample. 
The edge of the sample is still white signifying piling up of 
flux there, but the flux distribution at this time is much more uniform 
than it was earlier, and the current density is correspondingly much lower.
This is a feature caused by the short lifetime of currents of superconductors 
in the flux flow state.
The rightmost panels show the remanent state after the field has been 
increased to max $H_a=3$ and 
then back to $H_a=0$. The distributions are star-shaped, with the inner part 
of the sample has low current and contains a lot of trapped positive flux.
The flux is trapped due to a line with $H_z=0$ inside the sample, where the strong shielding 
currents flow with zero resistivity. The shielding from the currents at this line prevents 
the trapped flux from leaving the sample.

Let us return to the dimensional quantities to 
determine how easy it is to measure the effect of the dynamic barrier.
The most likely candidate for material are superconductors with
low intrinsic flux pinning and low first critical field. One such material is
MoGe thin-films.\cite{kubo88} With the values
$\mu_0H_{c2}=$3~T, $\rho_n=2\cdot 10^{-6}~\Omega$m,
$d=50~$nm, $w=2~$mm, and driving rate $\mu_0\dot
H_a=10~$T/s, we get $J_0=35~$A/m and $t_0=4.3~$ms.
The characteristic current density will 
thus be $J_0/d=6.9~10^{8}$A/m$^2$ and the magnetic field 
values will be of order $\mu_0J_0=0.043~$mT. 
In this case the dynamical barrier will be larger than the 
geometric barrier obtained 
Ref.~\onlinecite{zeldov94-prl}, which is of order 
$\mu_0H_p=\mu_0H_{c1}\sqrt{d/w}=0.01~$mT, with $\mu_0H_{c1}=2~mT$. Experimentally
it will thus be easy to distinguish the geometric barrier from the
dynamical barrier due to the ramp-rate dependency of the latter.  

\section{Summary}
The penetration of magnetic flux into superconducting films can be
delayed due to a dynamical barrier caused by the viscous motion of the
vortices. In this work we have studied this effect on a thin film
superconductor of square shape using numerical simulations. The point
that makes the dynamics interesting is that in transverse geometry,
the flux flow equations are non-linear due to the film self-field,
contrary to parallel geometry where they are linear. In small applied
magnetic field, the flux penetrates into the sample in a orderly
manner with a well-defined flux front, similar to the critical state,
but with absence of current discontinuity lines.  When the applied
field is changed there are fronts moving where the total magnetic
field is zero, and shielding currents flow without resistivity.  In
particular in the remanent state, such a front will prevent magnetic
flux from leaving the sample, so that the remanent state contains
trapped flux. The magnetization loop is hysteric with the
magnetization peak shifted from the zero position.  Numerical
estimates shows that the effect of the dynamical barrier should be
possible to measure on thin films of materials with low volume
pinning.  The effect is easily distinguished from other kinds of
barriers due to its dependence on the rate of change of the applied
field.

\acknowledgments
This work was financially supported by the Research Council of Norway.

\bibliography{superconductor}

\end{document}